\documentclass[
 aip,amsmath,amssymb,
 reprint]{revtex4-1}
\usepackage{epsf}  
\newcommand{\R}{\mathbb{R}}
\newcommand{\N}{\mathbb{N}}
\newcommand{\Z}{\mathbb{Z}}

\usepackage{amssymb}
\usepackage{graphics}
\usepackage{color}
\usepackage{graphicx}
\usepackage{dcolumn}
\usepackage{bm}
\usepackage[draft]{changes}
\usepackage{enumitem}

\usepackage{newtxtext}
\usepackage[colorlinks=true, pdfstartview=FitV, linkcolor=blue,
            citecolor=blue, urlcolor=blue]{hyperref} 

\usepackage[utf8]{inputenc}
\usepackage[T1]{fontenc}
\usepackage{amsthm}
\newtheorem{defi}{Definition}

\usepackage{sidecap}
\usepackage{grffile}
\usepackage{scalerel}

\usepackage{xcolor}

\usepackage[mathlines]{lineno}

\def\bea{\begin{equation} \begin{aligned}}
\def\eea{\end{aligned} \end{equation}}
\def\beas{\begin{equation*} \begin{aligned}}
\def\eeas{\end{aligned} \end{equation*}}
\def\bes{\begin{equation*}}
\def\ees{\end{equation*}}

\def\d{\mathrm{d}}

\def\be{\begin{equation}}
\def\ee{\end{equation}}

\newcommand{\Prob}{\mathbb{P}}

\definecolor{rred}{rgb}{0.7,0,0.1}
\definecolor{greenrb}{rgb}{0.2,0.6,0.2}

\begin{document}

\preprint{AIP/123-QED}

\title{Noise-driven Topological Changes in Chaotic Dynamics}
\author{Gisela D. Char\'o}
\email{gisela.charo@cima.fcen.uba.ar}
\affiliation{CONICET – Universidad de Buenos Aires. Centro de Investigaciones
del Mar y la Atmósfera (CIMA), C1428EGA  CABA, Argentina}
\affiliation{CNRS – IRD – CONICET – UBA. Institut Franco-Argentin d'\'Etudes sur le Climat et ses Impacts (IRL 3351 IFAECI), C1428EGA  CABA, Argentina.}

\author{Micka\"el D. Chekroun}
\affiliation{Weizmann Institute of Science, Rehovot 7610001, Israel.}

\author{Denisse Sciamarella}
\affiliation{CNRS – Centre National de la Recherche Scientifique, 75016 Paris, France.}%
\affiliation{CNRS – IRD – CONICET – UBA. Institut Franco-Argentin d'\'Etudes sur le Climat et ses Impacts (IRL 3351 IFAECI), C1428EGA  CABA, Argentina.}

\author{Michael Ghil}
\affiliation{Geosciences Department and Laboratoire de M\'et\'eorologie Dynamique (CNRS and IPSL), \'Ecole Normale Sup\'erieure and PSL University, 75231 Paris Cedex 05, France.}
\affiliation{Department of Atmospheric \& Oceanic Sciences, University of California, Los Angeles, CA 90095-1565, USA.}

\date{\today}

\begin{abstract}
\noindent
Noise modifies the behavior of chaotic systems in both quantitative and qualitative ways. To study these modifications, the present work compares the topological structure of the deterministic Lorenz (1963) attractor with its stochastically perturbed version. The deterministic attractor is well known to be ``strange'' but it is frozen in time. When driven by multiplicative noise, the Lorenz model's random attractor (LORA) evolves in time. Algebraic topology sheds light on the most striking effects involved in such an evolution. In order to examine the topological structure of the snapshots that approximate LORA, we use Branched Manifold Analysis through Homologies (BraMAH) — a technique originally introduced to characterize the topological structure of deterministically chaotic flows — which is being extended herein to nonlinear noise-driven systems.  The analysis is performed for a fixed realization of the driving noise at different time instants in time. The results suggest that LORA's evolution includes sharp transitions that appear as topological tipping points.
\end{abstract}

\maketitle

\begin{quotation}
H. Poincaré first described the way in which a dynamical system’s properties depend upon its topology \cite{Poincare1895}. The advantage of using topology instead of geometry or fractality to describe chaotic data lies in the fact that topology provides information about the stretching, folding, tearing and squeezing mechanisms that act in phase space to shape the flow. R. F. Williams introduced  the concept of branched manifold to characterize the surface of the Lorenz attractor \cite{williams1974expanding} and,  with J. Birman \cite{birman1983knotted}, used it to classify chaotic attractors in terms of the way in which their branches are knotted. This description goes beyond merely counting the branches: their organization and the presence of half-twists in some of them is  highly relevant. These features can be captured without dimensionality restrictions using homologies \cite{sciamarella1999topological}.  Here, we extend the usage of branched manifolds beyond the deterministic framework by investigating the evolution in time of the topological structure of the Lorenz Random Attractor (LORA) \cite{chekroun2011stochastic}. LORA's branched manifold is found to undergo abrupt changes at certain points in time, suggesting that the effects of noise on  chaotic dynamics can be addressed in this manner.  Topological tipping points can thus be defined as abrupt changes in the topology of a random attractor's branched manifold, and Branched Manifold Analysis through Homologies (BraMAH) emerges as a robust method that allows one to detect these fundamental changes.
\end{quotation}

\section{Introduction and Motivation}\label{sec:intro}

The use of topological concepts to describe complex physics goes back at least to Lord Kelvin, who proposed that atoms were knotted vortices swirling in the \texttt{\ae}ther, an invisible medium believed at that time to fill the surrounding space \cite{thomson1867ii}. Though incorrect, Kelvin’s vision stimulated active research in knot theory \cite{silver2006knot,thurston1979geometry} and eventually led to the awareness that knots and related tangled structures do actually form in various physical phenomena and can have a pivotal, albeit still poorly understood influence on turbulent fluid dynamics \cite{moffatt1981some,ricca1996topological}, quantum field theory \cite{witten1989quantum}, and magnetic fields \cite{glatzmaiers1995three,kedia2013tying}, to cite but a few areas of physics. In his seminal work, \citet{moffatt1969degree} showed that helicity measures the total knottedness and linkage of a flow and that it is an invariant in ideal, viscosity-lacking fluids, like liquid helium. In viscous fluids, helicity fluctuates, and knots can experience transformations or unravel \cite{moffatt2014helicity}.

In dissipative chaotic dynamical systems, applications of algebraic topology \cite{Poincare1895}, which studies algebraic invariants that classify topological spaces up to homeomorphism \cite{Hatcher2002}, have also attracted much attention over the past three decades \cite{mindlin1992topological,gilmore1998topological,muldoon1993topology}. Branched manifolds \cite{birman1983knotted, holmes1985knotted}, in particular, have played a central role in the characterization of the coarse-grained  features of chaotic dynamics.  The concept was anticipated in E. N. Lorenz’s famous 1963 paper \cite{lorenz1963lorenz}, where he remarks that the flow on the attractor's “surface” passes “back and forth from one spiral to the other without intersecting itself.” This surface is topologically equivalent to what is now called a branched manifold \cite{rossler2013chaos}.

\subsection{Branched Manifolds}\label{ssec:BMs}

A branched manifold has the property that each point in it has a neighborhood that is homoeomorphic to either a full ball or a half ball, thus allowing for the presence of boundaries, and therefore of branches \cite{muldoon1993topology}. Branched manifolds may thus have branches or not: they constitute a broader category, to which boundaryless manifolds belong as well. These objects provide a simple but powerful ``cartoon'' of the stretching and folding responsible for the creation of strange attractors, and are also referred to as a template or a knot holder. 

If the system is deterministic, the topology of its branched manifold is an invariant in phase space.  Metric invariants, such as  correlation dimension \cite{grassberger1983characterization}, and dynamic invariants, such as Lyapunov exponents \cite{wolf1985determining}, are very useful in describing and classifying dynamical systems. The topology of a branched manifold, though, provides more specific information on the mechanisms acting in phase space to shape the flow, and therefore on how to model the system’s dynamics \cite{gilmore1998topological}.
In the present paper, we extend the applications of branched manifolds to chaotic systems with time-dependent forcing, in particular to noise-driven ones.\\

\subsection{Noise-driven Chaos}\label{ssec:RDS} 

Interest in such nonautonomous and random dynamical systems (NDSs and RDSs) has been greatly stimulated recently --- among many other areas of application, as discussed in refs.~\cite{Cherubini.ea.2017,Faranda.ea.2017,sato2019anomalous} --- by the effects of anthropogenic forcing on the climate system, on the one hand, and by the multiscale character of the system's intrinsic variability, on the other \cite{Ghil.2019, ghil2019physics}. Several authors --- see ref.\cite[and further references therein]{berner2017stochastic} --- have addressed the latter problem by stochastic parameterizations of unresolved scales in high-order models, and more realistic low-order models have thus been obtained as well \cite{kondrashov2015data, chekroun2017emergence, santos2021reduced}.

With this motivation in mind, \citet{chekroun2011stochastic} studied interactions between noise and nonlinear effects from a qualitative dynamical viewpoint, through high-resolution numerical computations. The pullback attractors (PBAs) of RDS theory \cite{crauel1994attractors,arnold1998random} provide a natural framework for doing so and are the mathematically rigorous counterpart of the heuristically defined snapshot attractors of nonlinear physics \cite{romeiras1990multifractal}. Further  theoretical details on PBAs and RDSs appear in Appendices~\ref{app:PBA} and \ref{app:RDS}.

In the random, RDS setting, a PBA is computed by following an ensemble of trajectories, each driven by the same noise path.  This way, the well-known smoothing effect of noise due to a single very long integration of the stochastic system disappears in the pullback approach and the fractal structure of the chaotic dynamics is fully captured when the unperturbed nonlinear system's dynamics is itself chaotic.

PBAs thus provide a natural way to reveal the time-dependent stretching and folding caused by the interactions between noise and nonlinearities, when the unperturbed dynamics is chaotic\cite{chekroun2011stochastic}. In their study of the stochastically perturbed Lorenz \cite{lorenz1963lorenz} model's PBA, dubbed LORA, \citet{chekroun2011stochastic} pointed out that the stretch-and-fold mechanisms give rise to three distinct components of the flow in phase space: a smooth evolution tied to the Lorenz model's deterministic convection; a pervasive, local ``jiggling’’ due to the roughness of the  driving multiplicative noise; and sudden deformations of the random PBA's overall support. The static picture of the strange attractor generated by deterministic chaos \cite{Guck.Holm.1983, Ghil.Chil.1987} is thus replaced, in the presence of noise, by a dynamic version that is even stranger, and that we will refer to as {\it noise-driven chaos} in what follows.

Noise-driven chaos offers therewith a self-consistent framework for  extending the concepts  and tools of chaos topology towards more realistic situations in fluid dynamics, the climate sciences and elsewhere, to advance the understanding of complex nonlinear dynamics in the presence of noise. The purpose of this article is to examine noise-driven chaos through the lens provided by algebraic-topology tools.

\subsection{Algebraic topology tools}\label{ssec:TDA}

Topological data analysis studies the structure of large data sets by focusing on their connectivity \cite{chazal2017introduction,wasserman2018topological,murugan2019topology}. 
These methods are used more and more in numerous and diverse research areas, including image processing \cite{carlsson2008local}, the spread of social and biological contagions on networks \cite{taylor2015topological}, percolation theory \cite{speidel2018topological}, genomics and evolutionary dynamics \cite{ chan2013topology,rabadan2019}, and protein structure \cite{gameiro2015topological}.
 Even though topological data analysis refers broadly to all methods using notions of shape and connectivity to find structure in data \cite{wasserman2018topological},
the term is nowadays being used more narrowly to refer to a particular method called persistent homology (PH) \cite{edelsbrunner2008persistent, carlsson2014topological}.
This method has attained a great popularity in mathematical imaging and vision.

We study here LORA's changes in time by applying a topological data analysis methodology  to describe the topology of branched manifolds using homologies (BraMAH) \cite{sciamarella1999topological}. BraMAH was conceived three decades ago to replace the braids or knots made with reconstructed orbits, as done by Gilmore and co-workers \cite{gilmore1998topological} up to that point. Braids or knots dissolve into trivial objects beyond three dimensions and, besides, knotted or braided orbits are difficult to reconstruct from data sets that are not noise free. These shortcomings motivated the search for knot- or braidless and orbitless methods, as stated by Natiello and Solari \cite{natiello2007user, natiello2013braided}. 
	
Homologies, which had been used to analyze experimental dynamical systems involving boundaryless manifolds \cite{muldoon1993topology} provided an avenue for achieving this goal. 
The topological description provided by the Euler characteristic and Betti numbers associated with homology groups \cite{muldoon1993topology}, however, did not suffice to correctly identify a branched manifold from a point cloud. 
BraMAH went further to ensure the extraction of all the pertinent information, including torsions, branch crossings and weak boundaries \cite{charo_artana_sciamarella_2021}. 
Applied at the beginning to noisy experimental data \cite{sciamarella1999topological} and to the analysis of four-dimensional chaotic solutions of a Shilnikov type \cite{sciamarella2001unveiling}, BraMAH has been recently refined to handle 
data sets generated by Hamiltonian systems \cite{charo_artana_sciamarella_2021}, as well as by nonautonomous ones \cite{charo2020topology}.

In this work, we take a significant step forward, from the purely deterministic to the stochastic realm. Topology is no longer an invariant for RDSs, but it is still expected to encode the stretching and squeezing mechanisms shaping the flow in phase space. We define here an RDS's time-evolving branched manifold locally as an integer-dimensional set in phase space that provides a robust skeleton of the point cloud associated, at each instant, with the invariant measure supported by its random attractor; see Appendix~\ref{app:meas} for the theoretical underpinnings of the latter concepts. 

The point cloud in BraMAH is first decomposed into subsets of points, called cells, which are glued together into a cell complex that forms the coarse skeleton of the structure upon which the point cloud lies. In algebraic topology, a cell complex is a finite set of cells of different dimensions such that the interiors of the cells do not intersect \cite{kinsey1997topology}.
The BraMAH cell complex construction method is designed to approximate a point cloud lying on a branched manifold. This way, the cell complex's topological features unveil the topological signature of the branched manifold. The BraMAH methodology is briefly summarized and further refined in the next section and Appendix~\ref{app:bramah} provides further details.

The numerical results and their robustness are described in Sect.~\ref{sec:results}. The key results are:
\begin{itemize}[topsep=-0cm,itemsep=-1ex,leftmargin=0.5cm]
		\item [(a)] the BraMAH methodology captures LORA's time-evolving topological structure by constructing the branched manifold at each instant in time; 
		\item [(b)] LORA's branched manifold differs from the static one obtained for the deterministic Lorenz model's strange attractor; and 
		\item [(c)] the noise-driven model's otherwise gradually evolving branched manifold exhibits sharp transitions at discrete points in time.
\end{itemize}
These results are further discussed in Sect.~\ref{sec:conclude}.


\section{Branched Manifold Analysis through Homologies (BraMAH)} 
\label{sec:bramah}
Pursuing our goal of examining noise-driven chaos through the lens of algebraic-topology tools, we first provide some background on PH in general and on BraMAH in particular. 

\subsection{Background on PH}\label{ssec:back}

The PH method measures connectivity in terms of intersections between $\epsilon$-sized balls centered at each point in the point cloud. Connection rules to construct cell complexes can be defined in terms of these intersections as the value of $\epsilon$ grows from zero to some value which is related to the full size of the cloud. The set of rules establishing when and how to connect the points in the cell complex receives the name of filtration --- such as the \v{C}ech or Vietoris-Rips filtrations \cite{de2007coverage, edelsbrunner2008persistent} --- and $\epsilon$ is called the filtration parameter. For a given $\epsilon$-value, homologies can be used to identify holes (regions in the point cloud that are void of connections). 

Examining how the cell complexes change with $\epsilon$, holes can be described as having a lifetime, i.e., an $\epsilon$-range within which they ``persist''.  
The result of a PH study is often expressed using $\epsilon$ as the horizontal axis of a barcode, which furnishes a connectivity fingerprint of that particular point set. 
When applied to point clouds corresponding, for instance, to a data set generated by the equations of the Lorenz attractor \cite{maletic2016persistent}, one chooses a filtration method and obtains an $\epsilon$-dependent portrait of the point set. 
For a certain $\epsilon$-range, the PH study identifies the two holes of the butterfly's wings, along with many other holes, which are less persistent. But is any point cloud with two persistent holes in a certain filtration range identical to the Lorenz  attractor's branched manifold? Of course not. Counting the holes and pondering their persistence may be indicative of the attractor's identity but not conclusive. The information retrieved with a PH study contains valuable scale-dependent details within a point cloud, but the approach is not at all tailored to obtain the geometry-independent invariant of a deterministic dynamical system in phase space. 

Recently, \citet{strommen2021topological} have adapted PHs to yield a computationally tractable identification and classification of multiple weather regimes \cite{CDV.1979, Ghil.Chil.1987, Ghil.Rob.2002, Hannachi.ea.2017} on the basis of the topological structure of the atmospheric flow in a low-dimensional, high-variance subspace of a given data set. Given the large number of classification schemes --- based on phase space clustering and temporal-persistence criteria, among others \cite{Ghil.ea.S2S, S2S.2018} --- such a topological approach could provide additional insights into the still poorly understood mechanisms of the complex organization of the atmosphere's “regime diagram” \cite{Ghil.ea.2010.Hide}. The weather regime classification of ref. \cite{strommen2021topological} work resorts to bifiltrations in order to incorporate density as an extra filtration parameter into their PH study.

\begin{figure}[!ht]
	\includegraphics[width=\linewidth]{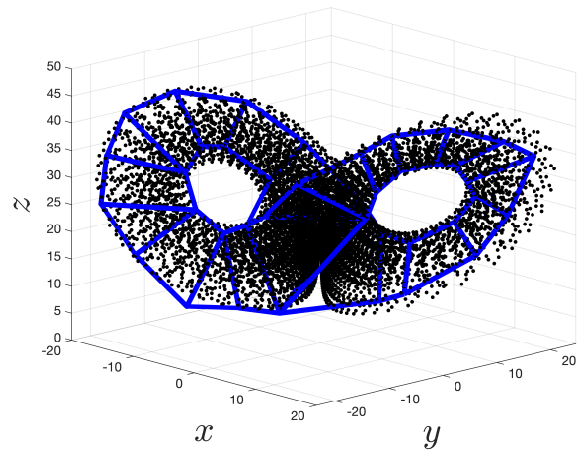}
	\caption{Point cloud (in black) juxtaposed on the cell complex  (blue borders) obtained by BraMAH for the deterministic \citet{lorenz1963lorenz} attractor, with $\sigma=0$ in Eq.~\eqref{eq_slm}.}
	\label{fig:det_lor}
	\vspace{-3ex}
\end{figure}

\begin{figure*}[!ht]
	\includegraphics[width=0.7\linewidth]{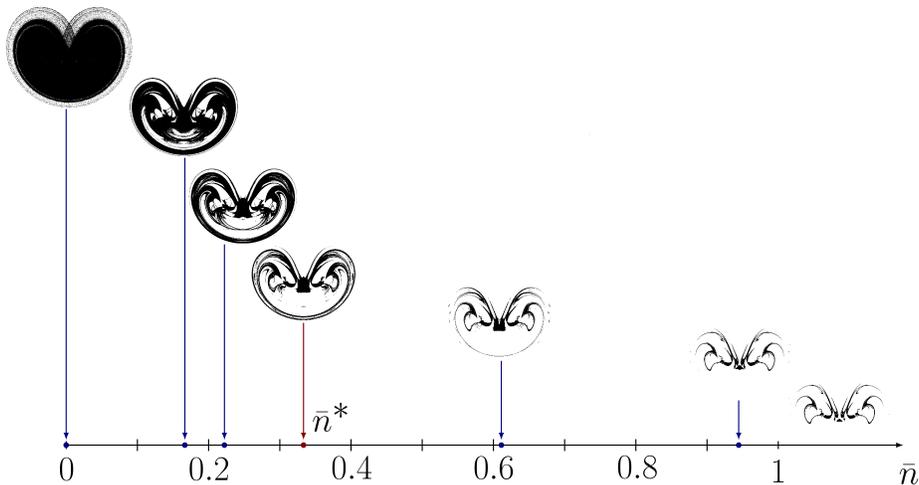}
	\caption{
		A projection onto the $(y,z)$-plane at time $t=40.27$ of the sieved point cloud obtained from the LORA simulation, for different normalized $\bar{n}$-values. The $\bar{n}$ value selected that unveils the most robust $1$-holes is $\bar{n}^* = 0.334$.}  
	\label{fig:nmax}
	\vspace{-3ex}
\end{figure*}

\subsection{The BraMAH Methodology}\label{ssec:method}

What does BraMAH do? BraMAH builds a cell complex that serves as an algebraic representation of the branched manifold
underlying a point cloud in phase space, and it computes the full set of topological properties that describe it. In other terms, BraMAH uses homologies as a means to reconstruct a branched manifold. The BraMAH methodology requires approximating the structure of the branched manifold that supports the point cloud by a cell complex,  as defined in algebraic topology \cite{kinsey1997topology}. 
The structure of this invariant manifold can be approximated with ``building blocks.'' In algebraic topology, these building blocks are Euclidean closed sets (segments, disks, etc.), called $n$-cells,  where $n \in \mathbb{N}_0$. A point is a $0$-cell, a line segment joining two points is a $1$-cell, a polygon is a $2$-cell and so forth. Constructing and assembling these cells according to the BraMAH rules detailed in Appendix~\ref{app:bramah} builds a cell complex that is the branched manifold's skeleton.

The topological structure of this cell complex is studied by using homology groups \cite{kinsey1997topology} and orientability chains \cite{sciamarella1999topological}. The homology groups $\mathrm{H}_k$ identify the $k$-dimensional holes ($k$-holes) of a topological space of dimension $n$,  where $k \in \mathbb{N}_0$ and $k\le n$. The group $\mathrm{H}_0$ identifies the connected components ($0$-holes), $\mathrm{H}_1$ the cycles ($1$-holes), $\mathrm{H}_2$ the cavities ($2$-holes), and if $k \ge 2$, $\mathrm{H}_k$ the hypercavities ($k$-holes). By construction, a BraMAH cell complex is uniformly oriented. The orientability chain indicates the number and location of torsions in the cell complex.
For further details of the BraMAH method, see Appendix~\ref{app:bramah}.  

As an example, we provide here a BraMAH analysis of the classical \citet{lorenz1963lorenz} attractor. 
The strange attractor and the BraMAH cell complex associated with its branched manifold is shown in Figure~\ref{fig:det_lor}. The cell complex presents one connected component, i.e., $\mathrm{H}_0 \sim \Z$, and  two homologically independent $1$-holes associated with each wing of the butterfly, i.e., $\mathrm{H}_1 \sim \Z^2$;  there are no torsions, i.e.,  $\mathrm{O}_1\sim \emptyset$, and no enclosed cavities, i.e., $\mathrm{H}_2\sim \emptyset$. 
The manner in which the two wings meet at the regular saddle on the $z$ axis can be assessed from the location of the $1$-holes, taking into account how they interweave without twisting.

In contrast to the PH approach, BraMAH's target is not counting holes and examining their persistence in terms of filtration rules that are established in terms of one or more free parameters. The manner in which BraMAH constructs a cell complex relies neither on $\epsilon$-sized balls around every point of the point cloud, nor on how such balls intersect.
 Knowing that the point cloud being analyzed lies on a branched manifold, one builds a single cell complex, and not a nested family of parameter-dependent cell complexes, as done in the PH approach. There is no analog of $\epsilon$ or of a filtration rule in BraMAH, since a branched manifold is not parameter-dependent in the PH filtration sense. This property of the dynamics generating the point cloud is utilized to produce a cell complex in which each cell represents a large subset of points that jointly approximate a locally Euclidean set, whose dimension respects the local dimension of the branched manifold. 
Some of the branches may have torsions that are important for the correct identification of an attractor; hence BraMAH also distinguishes, for instance, a standard strip from a M\"obius strip, or a M\"obius strip from a Klein bottle.  The method's specificities hinge on the cell complex construction process, as well as on the way it extracts the topological properties associated with the complexes so constructed.

\subsection{BraMAH and LORA}\label{ssec:BraLORA}

BraMAH has been applied so far to speech data \cite{sciamarella1999topological}, to slow-fast systems \cite{holmes1985knotted,deng1994constructing,sciamarella2001unveiling},  and to Lagrangian dynamics in fluid flows \cite{charo2020topology}. Here it is being applied for the first time to data sets generated by noise-driven chaos.
When evaluating a time-dependent structure like LORA, one has to ask whether its topology is time-dependent. In order to answer this question, we take two steps: (1) generate the point clouds that approximate LORA at successive time instants, called ``snapshots" \cite{chekroun2011stochastic, romeiras1990snap, Drotos.ea.2015}; and (2) compute the topological properties associated with each snapshot using BraMAH. 
 
In step (1), the point cloud approximating LORA is generated by solutions of the stochastic Lorenz model introduced by \citet{chekroun2011stochastic}. In \ref{eq_slm}, the equations of the classical \citet{lorenz1963lorenz} deterministic model are perturbed by a multiplicative noise in the It\^o sense \cite{arnold1998random}, with $W_t$ a Wiener process and $\sigma > 0$ the noise intensity:
\bea\label{eq_slm}
\vspace{-1ex}
&\d x=s(y-x)\d t+\sigma x {\mathrm d} W_t, \\
&\d y=(rx-y-xz)\d t+\sigma y {\mathrm d} W_t, \\
&\d z=(-bz+xy)\d t+\sigma z {\mathrm d} W_t; 
\vspace{-1ex}
\eea
here $r = 28$, $s = 10$, $b = 8/3$ are the standard parameter values for deterministically chaotic behavior. 

The time-dependent sample measures $\mu_t$ associated with the system ~\eqref{eq_slm} are probability measures for the population density of any ensemble of initial data driven by the same noise realization until time $t$, after removal of the transient behavior. Mathematically, these measures are of Sinai-Ruelle-Bowen (SRB) type \cite{eckmann1985ergodic, Young2002, Young2017}, i.e., they are supported by the foliation of unstable manifolds that structure the random attractor \cite{chekroun2011stochastic}.   

A numerical estimation $\hat{\mu_t}$ of such a measure can be computed at any time instant $t$ by a pullback approach, i.e.,~by letting a large set of $N_0$ initial points  $\{ \bm{x}_j (0) =  (x_j, y_j, z_j)(t=0): j=1, \ldots, N_0\}$ ``flow'' in phase space from the remote past until time $t$, for a fixed noise realization $\omega$. The convergence of the sample measure's approximation $\hat{\mu_t}= \hat{\mu_t}(N_0)$ is studied as the number $N_0$ of initial points increases; it is observed herein for $N_0 \simeq 10^8$.

Each point within a given cloud at time $t$ is mapped to a value of $\hat{\mu_t}$ that is obtained by averaging over a volume surrounding that point: higher or lower $\hat{\mu_t}$-values correspond to more or less populated regions of the random attractor. We  wish to characterize the topology of the point cloud's most populated regions, but also to ascertain this topology's robustness. 
In order to do this, a threshold for $\hat{\mu_t}$ must be selected, as discussed below.


\section{RESULTS}\label{sec:results}

\subsection{Methodological Results}\label{sec:methres}

Solutions of the stochastic Lorenz model (\ref{eq_slm}) generate data sets in the form of $3$-dimensional point clouds $ \{ \bm{x}_i \in \R^3 : 1 \le i \le N_0 \}$ with  $N_0=10^8$. In order to obtain the robust skeleton of the point cloud associated, at each instant, with the invariant measure supported by LORA, these point clouds are sieved using a threshold value $\bar{n}$ in the sample measure $\hat{\mu_t}$. This yields point clouds that reveal the most populated regions of the cloud. 

The $\bar{n}$-values are chosen within limits that guarantee that the connectivity of the point cloud is not lost, so that there is only one connected component. To this end, we define $\bar{n}_{\rm max}$ as the $\bar{n}$ value for which there is still one connected component, i.e., $\mathrm{H}_0 \sim \Z$, and normalize all $\bar{n}$ values by the value of $\bar{n}_{\rm max}$; for notational simplicity, we keep the symbol $\bar{n}$, though, for the normalized value as well.   

Figure~\ref{fig:nmax} shows how LORA at $t=40.27$ flakes off as the point cloud is sieved. We use the term ``sieve'' to avoid confusing this process with that of a filtration, as defined in PH.
Notice that when $\bar{n}>1$, the point cloud is broken into three sufficiently populated connected components ($\mathrm{H}_0 \sim \Z^3$).

In order to study LORA's changing topology at time $t$, a criterion must be adopted to select a value $\bar{n}^*_t$ of $\bar{n}$ that adequately represents the branched manifold for that value of $t$. For each $\bar{n}$-value, the BraMAH cell complex built from $\{ \bm{x}_i  ~:~ \hat{\mu_t}(\bm{x_i}) \ge \bar{n}  \}$ can be analyzed, yielding a set of 1-holes, among other results. As one can see in Figure~\ref{fig:nmax} for $t=40.27$, some of the 1-holes are robust as the value of $\bar{n}$ increases.  Based on this observation, we build a {\em robustness} plot, using $\bar{n}$ as the horizontal axis and the number of $H_1$ generators --- i.e., of 1-holes --- on the vertical axis, which pile up in the order of their appearance. 
Normalizing the lengths $\bar{\ell}$ of the bars against the longest bar $\bar{\ell}_{\rm max}$ of the plot, we define the most robust 1-holes as those for which  $\bar{\ell}/\bar{\ell}_{\rm max} \geq 0.5$. With this criterion of robustness in mind, we finally require $\bar{n}^*_t$ at each time $t$ to be a value of $\bar{n}$ at which the $H_1$ generators of the BraMAH cell complex coincide with the most robust $1$-holes.

Note that the density here is not a filtration parameter in the PH sense, i.e. the value of $\bar{n}$ is not involved in the rules used to construct the BraMAH cell complex. For this reason, the robustness plot should not be read as a PH barcode: at each $\bar{n}$-value, BraMAH uses a point cloud that differs in its size  and elements 
to construct the cell complex, whereas in PH calculations single point cloud is under consideration and the horizontal axis sweeps parameter values that regulate the construction of nested cell complexes for the same point cloud. 
\begin{figure*}
\includegraphics[width=\linewidth]{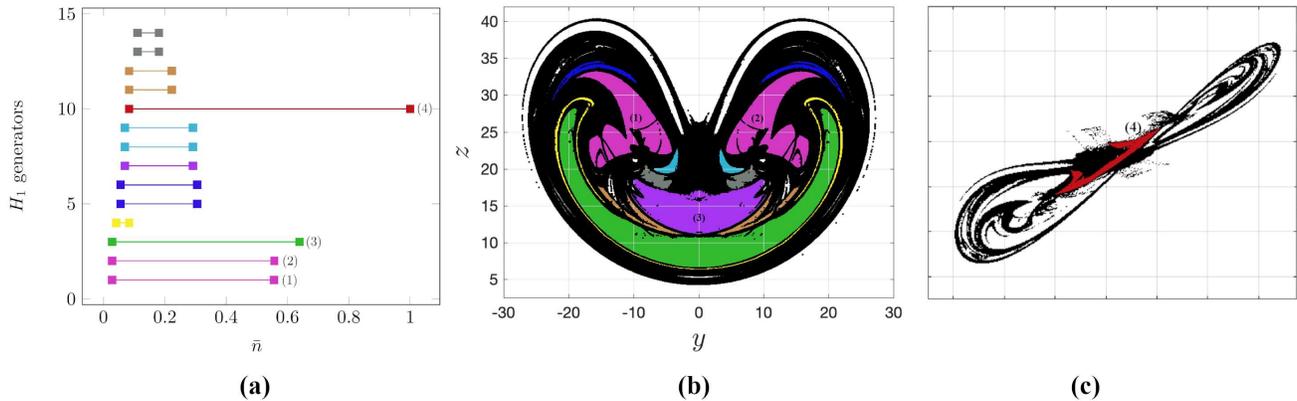}
\caption{Robustness characteristics of the BraMAH computation. (a) Robustness plot for LORA at $t=40.27$ with $\sigma = 0.3$. Fourteen 1-holes are represented and colored. The most robust holes are labeled with numbers (1)--(4). Plot (b) is an $(y,z)$ projection of the sieved point cloud ($\bar{n}=0.167$) where holes labeled (1)--(3) can be identified. Plot (c) is a projection onto the plane $-2.25x-20y+6z=0$ of the same point cloud where hole labeled (4) can be visualized. The colors of the holes correspond to those of the bars in (a).}
\label{fig:persistence}
\vspace{-4ex}
\end{figure*}

Figure~\ref{fig:persistence}(a) shows the robustness plot for $t=40.27$ with noise intensity $\sigma = 0.3$. Fourteen $H_1$ generators (1-holes) are represented and colored. The most robust holes are labeled with numbers (1)--(4). In panel (b), we show an $(y,z)$ projection of the sieved point cloud with $\bar{n}=0.167$, in which holes labeled (1)--(3) can be identified. In (c), we show a projection onto the plane $-2.25x-20y+6z=0$ of the same point cloud where hole labeled (4) can be visualized. The colors of the holes correspond to those of the bars in (a).

\subsection{Topological Results}\label{ssec:topo}

Following the criteria formulated in Sects.~\ref{sec:bramah} and \ref{sec:methres} and illustrated in Figs.~\ref{fig:nmax} and \ref{fig:persistence}, we analyze the topology of three successive LORA snapshots for a given, fixed noise intensity of $\sigma = 0.3$.  These three points in time --- $t = 40.09, 40.18$ and $40.27$ --- have been chosen visually from the LORA video provided  in the Supplementary Material of \citet{chekroun2011stochastic}. The video clearly suggests a gradual, smooth evolution of the invariant measure, except for sudden changes at discrete times. Two such points in time appear to lie in the open intervals $(40.09 , 40.18)$ and $(40.18, 40.27)$, both of which are quite short with respect to the characteristic nondimensional time of $1.0$ for the deterministic Lorenz model \cite{lorenz1963lorenz}.
	
At these three points, we find that the local dimension of the branched manifold is the same as in the deterministic case:  locally, LORA remains a $2$-branched manifold and no torsions are ever observed, i.e., $\mathrm{O}_1\sim \emptyset$. Noise adds neither connected components nor cavities to the manifold, so that the groups $\mathrm{H}_0 \sim \Z$ and $\mathrm{H}_2\sim \emptyset$ are the same as in Figure~\ref{fig:det_lor}. The topological distinctions between LORA and the classical  \citet{lorenz1963lorenz} attractor are thus solely present in $\mathrm{H}_1$. One notices immediately, though, that the number of $1$-holes changes from one snapshot to the next, with holes created or destroyed by the noise in the course of time.

\begin{figure*}[!ht]
	\includegraphics[width=\linewidth,height=0.5\textwidth]{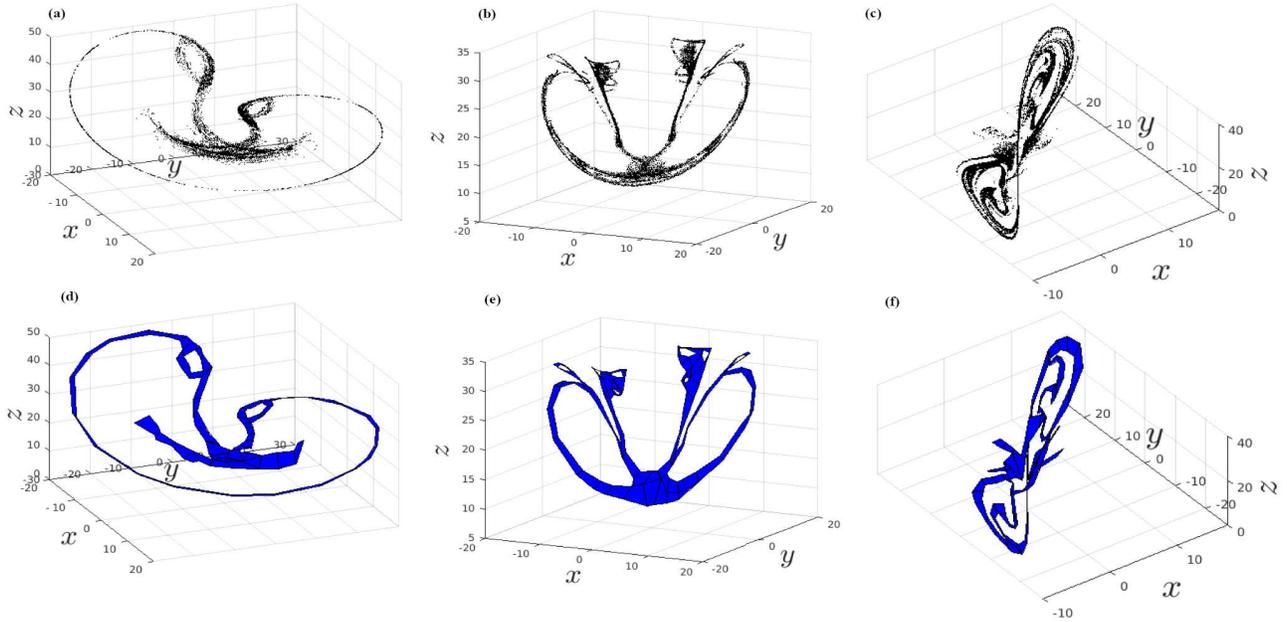}
	\caption{Three LORA snapshots with $\sigma=0.3$ and $N_0 = 10^{8}$: 
		sieved point clouds (a)--(c) and cell complexes (d)--(f). (a,d) $t=40.09$, $\bar{n}^{*}=0.5$, $\mathrm{H}_1 \sim \Z^{3}$; (b,e) $t=40.18$, $\bar{n}^{*}=0.9$, $\mathrm{H}_1 \sim \Z^{10}$; and (c,f) $t=40.27$, $\bar{n}^* = 0.334$, $\mathrm{H}_1 \sim \Z^{4}$.}
	\label{fig:LORA_complex}
	\vspace{-3ex}
\end{figure*}

This situation is illustrated in Fig.~\ref{fig:LORA_complex} for the three successive snapshots, at $t=40.09, 40.18$ and $40.27$; for each snapshot, we use the value of $\bar{n}^*$ specified in the caption, which leads to LORA's branched manifold at time $t$. The sieved point clouds are given in the 3 top panels, while the corresponding BraMAH cell complexes are presented in the 3 bottom panels. The branched manifold's structure undergoes large changes not only in the number of $1$-holes but also in the organization of the branches from one snapshot to the next.

The BraMAH analysis herein thus fully supports the intuition provided by the LORA video of \citet{chekroun2011stochastic} on sudden topological changes associated with those visible in the invariant measure. Further work clearly needs to complete the present analysis in terms of narrowing down even further the exact location of the topological tipping points and being more exhaustive in cataloguing the possible changes in the branched manifolds, for LORA and for other stochastically perturbed chaotic systems.


\section{Concluding Remarks}\label{sec:conclude}
We have topologically analyzed here a paradigmatic random attractor \cite{chekroun2011stochastic} associated with the  \citet{lorenz1963lorenz} convection model.  To accomplish this, we have extended the deterministic concept of branched manifold \cite{williams1974expanding}, by defining it locally as an integer-dimensional set in phase space that robustly supports the point cloud associated with the system's invariant measure at each instant $t$. This new definition of branched manifold is used to perform what is --- to the best of the authors' knowledge --- the first topological analysis of a time-evolving chaotic attractor.
 
 Our topological analysis is based on a topological data analysis methodology called BraMAH \cite{sciamarella1999topological, sciamarella2001unveiling, charo2020topology, charo_artana_sciamarella_2021}, originally designed to study the topological structure of chaotic data in a deterministic setting \cite{sciamarella1999topological}. This homology computation tool leads from a point cloud to the full set of topological features that 
characterize a branched manifold. BraMAH differs from other generic topological data analysis methods by its using explicitly the fact that the point cloud under study lies on a branched manifold, and by detecting relevant features such as branch torsions or weak boundaries, if they exist. 

 In the presence of noise-driven chaos, the branched manifold is time-evolving and must therefore be analyzed time-wise, i.e., using point clouds that approximate the random attractor by successive snapshots.
The branched manifolds so constructed by BraMAH  provide topological portraits at different stages in the life of LORA, the Lorenz model's random attractor.

Our study shows a marked difference between the deterministic case and the noise-driven one. The stochastically perturbed system's random attractor, dubbed LORA, presents a much richer structure than the deterministic strange attractor, with a topology that also changes drastically in time.  The framework introduced in this article to characterize such changes in topological features appears to hold promise for the understanding of topological tipping points  in general.

A fairly straightforward BraMAH application to the climate sciences might clarify the following quandary of so-called subseasonal-to-seasonal prediction \cite{S2S.2018}. The quandary deals with the role of intermittent vs. oscillatory low-frequency variability  in the atmosphere \cite{Ghil.Rob.2002, Ghil.ea.S2S}. Low-frequency variability refers to phenomena whose characteristic time of 10--100~days exceeds that of typical midlatitude storms of 5--7~days;  see, for instance, \citet[Ch.~6]{Ghil.Chil.1987}.

Such phenomena include the so-called blocking of the westerlies and intraseasonal oscillations with periodicities of 40--50~days; the former is intermittent, or particle-like, the latter oscillatory or wave-like \cite{ghil2019physics, Ghil.Rob.2002}. The quandary is which type of phenomenon dominates low-frequency variability and thus could contribute most to subseasonal-to-seasonal prediction, cf.~\citet[Fig.~1]{Ghil.2020}.

Blocking has been recently studied in \citet{Lucarini.Gritsun.2020} by the unstable periodic orbits methodology of \citet{gilmore1998topological}. It would appear that the BraMAH methodology proposed herein could address, more generally and efficiently, the waves-vs.-particles quandary of low-frequency variability and how it might be affected by global change \cite{Ghil.2020}.

More broadly, tipping points have been given a precise definition in the climate sciences as a generalization to NDSs and RDSs of classical bifurcations in autonomous systems \cite{Ashwin.ea.2012, Ghil.2019} and they are being actively pursued \cite{Tel.ea.2020, Pierini.ea.2016, Pierini.ea.2018, CGN.2018, Pier.Ghil.2021, Vann.ea.2021}. Topological tipping points seem to be a further generalization of the concept that could help us apprehend sudden and drastic changes in time of model behavior, as well as drastic changes due to mean forcing intensity.  

The gradual change of atmospheric concentrations of greenhouse gases and aerosols may continue to modify global temperatures in a fairly smooth way \cite{IPCC.AR4.2007}, as the case has been so far, although it might also lead to dynamical tipping points, as suggested by the above-cited papers. On the other hand, the intrinsic noise associated with cloud processes on small space and time scales affects the entire climate system through their interaction with dynamic and radiative processes on larger scales and may lead to hitherto unsuspected topological tipping points \cite{ghil2019physics}. 

Topological tipping points seem to be a further generalization  of the concept that could help us apprehend sudden and drastic changes in time of model behavior, as well as drastic changes due to mean forcing intensity, whether deterministic or stochastic.

We have concentrated throughout much of this paper on problems related to the climate sciences, as had E.N. Lorenz in his pioneering paper \cite{lorenz1963lorenz}. With all due modesty, it is not 
unlikely — considering the great generality of topological  methods \cite{Poincare1895,gilmore1998topological} — to expect the results obtained herein to have some applicability to other areas of the physical, life and socio-economic sciences.

\begin{acknowledgments}
We thank many colleagues who asked good questions at several talks and posters given on this work; see, for instance, refs\cite{charo:hal-02371641, charo2020topological}. It is a pleasure to thank three anonymous reviewers whose warm reception of this paper's originally submitted version and constructive suggestions greatly improved the final version.
This work is supported by the French National program LEFE-NOISE (Les Enveloppes Fluides et l’Environnement) and by the CLIMAT-AMSUD 21-CLIMAT-05 project (D.S.). G.D.C. gratefully acknowledges her postdoctoral scholarship from CONICET and wishes to thank Juan Ruiz and the CIMA computing staff for support with the computations. This work was also partially supported by the European Research Council (ERC) under the European Union’s Horizon 2020 research and innovation program (grant agreement No.~810370), by a Ben May Center grant for theoretical and/or computational research and by the Israeli Council for Higher Education (CHE) via the Weizmann Data Science Research Center (M.D.C). The article is TiPES contribution \#46; this project has received funding from the  European Union's Horizon 2020 research and innovation program under grant agreement No.~820970 (M.G.).
\end{acknowledgments}

 \section*{Data Availability}
 
 The data that support the findings of this study are available from the corresponding author upon request.

\vspace{-3ex}

\appendix
\section{PULLBACK ATTRACTORS}\label{app:PBA}
We summarize herein some pertinent facts on nonautonomous systems of ordinary differential equations, 
\begin{equation}
\dot{\bm{x}}=\bm{F}(t,\bm{x}), \; t \in  \mathbb{R},
\label{eq:nonaut}
\end{equation}
considered in the framework of nonautonomous dynamical systems (NDSs); $F$ denotes a smooth time-dependent vector field that governs the time evolution of the state $x$ in a phase space $X$, taken for simplicity to be the Euclidean space $\mathbb{R}^N$.

Once existence and uniqueness are guaranteed, one can assign to this NDS a solution map $\Phi(t,s)$, which provides a two-time description of the motion: the time $s$ when the system was initialized, and the time $t\geq s$ of the system's current state. Thus 
$$x(t) = \Phi(t,s) x_0$$ 
denotes the solution of Eq.~\eqref{eq:nonaut} at time $t$, when initialized at  $x(t) =x_0$ for time $t=s$. In the autonomous case, only the time interval $t' = t - s$ separating $s$ and $t$ matters and $\Phi(t,s)$ reduces to a standard flow $\Phi(t')$. 

In the case of forced and dissipative systems, such as the climate system \cite{Ghil.Chil.1987, ghil2019physics}, one can define a collection of subsets called a \textit{pullback attractor} (PBA) \cite{arnold1998random,caraballo2006pullback, 
ghil2008climate, chekroun2011stochastic, CGN.2018}. 
\begin{defi}
A PBA is a family 
$ \{\mathcal{A}(t)\}_{t\in \mathbb{R}}$, where $\mathcal{A}(t)$ is a compact subset of $X$ at each time $t$.
For each $ t \in \R$, this family has two fundamental  properties:
\begin{itemize}
\item [(1)] {\bf Invariance:} $\Phi(t,s) \mathcal{A}(s) = \mathcal{A}(t)$, for all $t \geq s$,  and
\item [(2)]  {\bf Pullback attraction:} For  any nonempty subset $B$ of  $X$, 
$$\displaystyle \lim_{s \to{-}\infty}{d_X \left(\Phi(t,s)B,\mathcal{A}(t)\right)}=0,$$ 
where $d_X$ is the Hausdorff semi-distance in $X$.
\end{itemize} 
\end{defi}
According to (1) and (2) above, the family $\mathcal{A}(t)$ is invariant under the system's dynamics and it attracts at each time $t$ all compact initial subsets $B$ from the remote past; see also \citet[Fig.~A.1,]{ghil2008climate} for a simple illustration.


\section{RANDOM DYNAMICAL SYSTEMS (RDSs)}\label{app:RDS}

In physical systems, such as those encountered in the climate sciences, random time-dependent forcing is often present \cite{ghil2019physics}. When that is so, it becomes necessary to model this type of system using stochastic differential equations (SDEs) \cite{arnold1998random}.

In the theory of RDSs, random PBAs are known as {\it random attractors} and they can be constructed in an extended phase space composed of the phase space $X$ and a probability space associated with the paths of the driving noise. A probability space $(\Omega,\mathcal{F},\mathbb{P})$ is a three-tuple, where $\Omega$ is the sample space; $\mathcal{F}$ is the event space, formulated as a $\sigma$-algebra; and $\mathbb{P}$ is a probability measure on $\mathcal{F}$; see \cite[Appendix A]{arnold1998random}.

The probability space is then endowed with a time-dependent shift  $\theta_t$. In the case of a stochastic-dynamic system driven by a Wiener process, as is the case here, this shift is defined on $\Omega$ according to $W_s(\theta_t \omega)=W_{t+s}(\omega)-W_s(\omega)$ \cite{arnold1998random}. With this mapping $\theta_t$ in hand, the noise realization $\omega$ evolves in time, and one can define a cocycle to describe the evolution of the state $x$.

A mapping $\varphi:\R \times \Omega\times X \to X$ has the cocycle property when $\varphi(t,\omega)= \varphi(t,\omega,\cdot):X \to X$ satisfies the following conditions \cite{arnold1998random}:
\begin{itemize}
    \item [(i)] $\varphi(0, \omega) x=x$,  for  all $x \in X$ and $\omega \in \Omega$, and
     \item [(ii)] $\varphi(t + s, \omega) = \varphi(t, \theta_s(\omega))\circ \varphi(s,\omega)$, for all $s,t \in \R$ and 
     $\omega \in \Omega$, where $\circ$ denotes the composition operation for mappings of $X$.
\end{itemize}
Property (i) just sets the initial state of the cocycle, while property (ii) states that a cocycle is an expression of the existence and uniqueness of solutions, in the sense that going from a copy of $X$ at time $0$ to one at time $s$ and from there on to one at time $t+s$ is the same as going directly from $0$ to $t+s$; see also \citet[Fig.~A.2]{ghil2008climate}. This cocycle property is satisfied for a broad class of SDEs like those of interest here; see \citet{arnold1998random}.

Mathematically, given an SDE with the right properties, the probability space $(\Omega,\mathcal{F},\mathbb{P})$ equipped with the collections of shifts $\theta = \{\theta_t\}_{t\in \R}$, and its associated cocycle $\varphi$, form what is called an RDS $(\varphi,\theta)$, also called sometimes an RDS $\varphi$ over $\theta$.

The evolution of a stochastic-dynamic system can thus be modeled by an RDS, while its associated random attractor $\{{\mathcal{A}(t; \omega)}\}_{t\in \mathbb{R}}$ provides the natural extension of a PBA (as defined in Appendix~\ref{app:PBA} above) to the random setting in which each individual PBA depends on the specific noise realization $\omega \in \Omega$. 
 The resulting family $\bigcup_{\omega \in \Omega}{\mathcal{A} (t;\omega)}$ of random compact sets  provides a complete description of all the system's possible states that are likely to be observed at time $t$.
 

 \section{INVARIANT MEASURES}\label{app:meas}
 
A number of interesting properties follow from the fact that the RDS $(\theta,\varphi)$ has a random attractor. One of these is the existence of invariant measures of $(\theta,\varphi)$, in the sense of RDS theory. In this Appendix, we briefly clarify this notion and discuss the properties of these measures.

To do so, recall that any RDS  $(\theta,\varphi)$ generates a skew-product semiflow $\{\Theta(t)\}_{t\geq 0}$ on $\Omega \times X$ by the formula
\be
\Theta(t)(\omega,x)=(\theta_t \omega, \varphi(t,\omega)x), \; t\geq 0.
\ee
The cocycle property for $\varphi$ is equivalent to the semigroup property for $\Theta(t)$, namely $\Theta(t+s)=\Theta(t)\Theta(s)$.
In what follows we denote by $\mathcal{B}$ the $\sigma$-algebra of Borel sets in $X$; see \citet{arnold1998random}.
We have then the following definition.

 \begin{defi}\label{Def_IM}
 Given an  RDS $(\varphi,\theta)$, a probability measure $\mu$ on $\left(\Omega \times X,\mathcal{F} \times \mathcal{B} \right)$ is called an invariant measure for $\varphi$ if it satisfies:
 \begin{itemize}
     \item [(i)]$\Theta(t)\mu=\mu$, for all $t \in \R$.
     \item [(ii)] The basic probability measure $\mathbb{P}$ is the marginal on $(\Omega,\mathcal{F})$ of $\mu$, 
     i.e.~$\mu(E\times X) =\mathbb{P}(E)$ for any $E \in \mathcal{F}$. 
 \end{itemize}
 \end{defi}
 
It is known that any probability measure $\mu$ on $\left(\Omega \times X,\,\mathcal{F} \times \mathcal{B} \right)$ possesses a {\it disintegration} or factorization \cite{arnold1998random}, given by a function $(\omega,B) \mapsto \mu_{\omega} (B)$ from $\Omega\times \mathcal{B}$ into the interval $[0,1]$ such that:
 
\begin{itemize}
 \item [(i)] For any $B \in \mathcal{B}$, $\mu_{\omega}$ is $\mathcal{F}$-measurable;
 \item [(ii)] there exists a measurable set $\Omega'$ in $\Omega$ such that $\Prob(\Omega')=1$ and $ \mu_{\omega}$ is a probability measure on $(X,\mathcal{B})$ for all $\omega$ in $\Omega'$; and
  \item [(iii)] for all $f$ in $L^1_\mu(\Omega\times X)$ we have  
  \be
 \displaystyle \int_{\Omega\times X} f(\omega, x) \mu(\d \omega, \d x) = \int_{\Omega} \bigg(\int_X  f(\omega,x) \mu_{\omega} (\d x)\bigg) \Prob(\d \omega).
  \ee
\end{itemize}
The disintegration $\mu_\omega$ is unique $\Prob$-almost surely and it is also called a {\em sample measure}; see \citet{Young2002}. The invariance property  (i) of Definition \ref{Def_IM} 
translates into $\varphi(t,\omega)\mu_{\omega}=\mu_{\theta_t \omega}$, in terms of sample measures. 
 
When an RDS $(\theta,\varphi)$ possesses a random compact attractor, then it supports every invariant measure, i.e.~$\mu_{\omega} (\mathcal{A}(\omega)) = 1$ for almost all $\omega$ in $\Omega$. In this case, the sample measure possesses a useful interpretation. To understand it, recall that, roughly speaking, the random attractor $\mathcal{A}(t; \omega)$ determines the portions of the phase space $X$ onto which any bounded set $B$ is mapped at time $t$, when $B$ is propagated by the cocycle $\varphi$ from a remote past, for a given noise realization $\omega$. The sample measure $\mu_{\theta_t\omega}$ supported by the random attractor $\mathcal{A}(t; \omega)$ provides, therewith, the spatio-temporal probability distributions of the portions of the phase space $X$ occupied by the RDS, at time $t$ and for the noise realization $\omega$.

For a given stochastic-dynamic system, the set of possible invariant measures is rather large. This raises the question whether a particular class of invariant measures is ``naturally chosen" by the dynamics. {\em Physical measures} are sample measures of special interest in this respect \cite{chekroun2011stochastic, ghil2019physics}. A probability measure $\mu$ is physical if, for any continuous observable $\psi: X \to \R$, the time average equals the ensemble average for almost all initial data $\bm{x}_0$ that lie in a  Lebesgue-positive set $B_{\mu}$, called the basin of attraction of $\mu$; see \citet[Eq.~(5)]{chekroun2011stochastic}. 

Sinai-Ruelle-Bowen (SRB) measures \cite{eckmann1985ergodic, Young2002, Young2017} form a closely related class of probability measures. A probability measure $\mu$ is an SRB measure if its conditional measures on the unstable manifolds are absolutely continuous with respect to Lebesgue measure. For many dynamical systems, the class of physical measures coincides with that of SRB measures; small differences may exist, however, between the two concepts for certain systems. Regarding the stochastic Lorenz system \eqref{eq_slm}, \citet{chekroun2011stochastic} have rigorously shown in their Appendix~C that LORA supports a random SRB measure  when the Kolmogorov operator  associated with \eqref{eq_slm} is hypoelliptic and its leading Lyapunov exponent is positive. Both conditions are met for the parameter values and stochastic forcing used in Eq.~\eqref{eq_slm}. Strong numerical evidence provided in ref.\cite{chekroun2011stochastic} suggests that the measure to which the point clouds used herein converge must be physical, too.

Note that the existence of an SRB measure $\mu$ does not guarantee its uniqueness, and that two such measures $\mu \neq \nu$ may also have different basins of attraction, $B_{\mu} \neq B_{\nu}$. The extensive numerical calculations in ref. \cite{chekroun2011stochastic} and herein have given no indication, though, of such nonuniqueness. Still, the uniqueness of LORA's random SRB measure or of its physical measure has not been proven rigorously, to the best of the authors' knowledge. 

\section{BraMAH  Implementation}\label{app:bramah}

BraMAH computes the topology of point clouds whose points are locally distributed on a branched manifold. An $m$-manifold is a topological space with the property that each point has a neighborhood that is homeomorphic to either a full $m$-ball or a half $m$-ball, with $m$ in $\N$ \cite{muldoon1993topology}. In nonlinear dynamics, an $m$-branched manifold is a mathematical object embedded in a phase space of dimension $d$ ($m \le d$) that is a manifold everywhere but at the tear points of the branches \cite{gilmore1998topological}. We refer to $m$ as the local dimension of the branched manifold.

Subsets of points of such a point cloud can be locally approximated by $m$-disks. This decomposition into point subsets or ``patches" is called ``patch decomposition" \cite{sciamarella1999topological}. A BraMAH complex is a cell complex constructed so that each patch is associated with a cell in the complex. 

To define cell complexes in general, we start with a single cell. A cell of dimension $k$ ($k$  in $\mathbb{Z}^+$) or $k$-cell 
is a set that can be mapped through a homeomorphism into the interior of a $k$-disk, so that the boundaries of the image are divided into a finite number of lower-dimensional cells, called faces. A cell complex $\mathbb{K}$ is a finite set of cells, such that (a) their faces are also elements of the complex; and (b) the interiors of two cells never intersect. $\mathbb{K}$ is said to be an $h$-complex, or to have dimension $h$, if its highest dimensional cell is an $h$-cell. 

How does BraMAH construct cells from patches? One first partitions the point cloud $\{\bm{x_i} \in \R^d : 1 \le i \le N \}$ into overlapping patches that are homeomorphic to the interior of an $m$-disk. A patch  $\{\bm{x_i}=(x_{i,1}, x_{i,2}, \ldots , x_{i,d} ), i=1, \ldots ,N_c \}$ is built around a center  $\bm{x_0}=(x_{0,1}, x_{0,2}, \ldots , x_{0,d} )$ by searching for the largest number $N_c$ of points around $\bm{x_0}$ that constitute approximations to a Euclidean set of dimension $m \le d$. In order to calculate $N_c$, candidate sets $\{\bm{x_i}, i=1, \cdots ,n_c \}$ are computed with its elements sorted by distance from $\bm{x_0}$, and with $N_{\rm{min}} \le n_c \le N_{\rm{max}}$. We use $N_{\rm{min}}$ typically $2\%$ the total size of the cloud, and $N_{\rm{max}}$ an order of magnitude higher than $N_{\rm{min}}$.

Given a candidate set, the points $\bm{x_i}$ within the ball with center $\bm{x_0}$ and radius $r$ are represented by the neighborhood matrix $X \in \R^{n_c \times d}$:
\be\label{eq:neighbor}
	X_{i,j} = \displaystyle{\frac{1}{n_c^{1/2}}(x_{i,j}-x_{0,j})}.
\ee
A local coordinate system centered at $\bm{x_0}$ is provided by the singular vectors of $X$, with the singular values describing the distribution of the points inside the ball centered at $\bm{x_0}$. For a patch that is approximately lying on an $m$-disk in $\R^d$, the local singular spectrum of $X$ has $m$ singular values that scale linearly with $n_c$ as $r$ is increased. This property holds while the effects of the manifold's curvature are negligible \cite{muldoon1993topology,broomhead1987topological}. The remaining $(d - m)$ singular values, which measure the deviation from the tangent space, will scale as $r^{\ell}$ with $\ell \ge 2$. 

Using this rule, the $m$ relevant singular values and vectors that span the tangent space approximating the patch under consideration can be identified. The value of $N_c$ is obtained when the $m$ relevant singular values --- as functions of $n_c$ subject to $N_{\rm{min}}< n_c <N_{\rm{max}}$ --- exhibit the best linear regression coefficient. 

This procedure is applied repetitively until every point of the point cloud belongs to at least one patch. The patch axes are chosen so as to favor patch overlapping, in order to keep track of the gluing prescriptions between them. Each patch is transformed into a cell by using convex hulls, and the singular vectors are used to orient each cell so that neighboring cells have the same orientation, thus resulting in the complex having a uniform orientation. A BraMAH complex is a cell complex that results from applying these steps to an $d$-dimensional point cloud. 

The next step in the BraMAH methodology is to compute the topological properties of the cell complex obtained from the algorithm above. For any given complex, a $k$-chain is a linear combination of $k$-cells with integer coefficients. The algebra of these chains allows one to describe the connectivity of the cells at each $k$-level. A $k$-hole is a closed chain, called a $k$-cycle, that is not the border of any higher-dimensional cell.  

Our approach uses the labeled list of $0$-cells of the BraMAH complex in order to build a boundary matrix and extracts the linearly independent rows of it. Then, it computes the null spaces of the transpose of the boundary matrix and expresses the $k$-borders in terms of the $k$-cycles to determine which $k$-cycles are homological to others. The $k$-cycles that are homologically independent are appended to $\mathrm{H}_k$. The integer multiples found in the chain that sums up all the $k$-borders of the complex are used to form the orientability chain, and the orientability chain is in turn used to compute and locate torsions, as well as weak boundaries. 

The results of this final phase of the method are the homology groups of the BraMAH complex expressed in terms of their generators $\{\mathrm{H}_k: k=0, \ldots, h\}$, spelled out in terms of the $0$-cells.
The advantages of the BraMAH methodology over other homology computation methods are the following:\\ 
\begin{itemize}[topsep=-0.3cm,itemsep=0ex,leftmargin=0.5cm]

	\item [(i)] A BraMAH cell complex is uniformly oriented --- thus allowing for the computation of orientability properties --- and it has a dimension $h$ that agrees with the local dimension $m$ of the branched manifold; the latter is computed locally using the neighborhood matrix $X$ so that no cells with unnecessarily high dimensions are created. 
	
	\item [(ii)] This cell complex is non-simplicial by construction, i.e., the number of sides of a cell is free, and each $h$-cell is constructed from a large subset of points; doing so yields a complex with a number of $h$-cells that is significantly lower than the number of points in the original point cloud, while PH-type methods --- e.g., the Vietoris-Rips complex --- tend to produce complexes with a number of cells that is vastly larger \cite{de2007coverage, edelsbrunner2008persistent}. 
	
	\item [(iii)] The $0$-cells of the BraMAH complex are associated with a set of points whose coordinates are identified in the original point cloud; these coordinate values can be used to embed the complex in phase space and to ascertain the mutual organization of the branched manifold's branches. 
	
	\item [(iv)] The $k$-holes and orientability chains of the complex can be superimposed on the point cloud or --- if $d \ge 4$, on a projection of the point cloud --- allowing one to identify the branches, as well as the torsions or twists along them.	

\end{itemize}
\vspace{0.3cm}

\noindent
As discussed in Section \ref{sec:bramah}, BraMAH is a tpological data analysis approach that does not fall into the category of PH methods. The PH approach focuses on how connectivity properties of a point cloud change with respect to a parameter $\epsilon$ that is involved in the filtration rule, i.e. in the rule used to build the cell complex. PH cell complexes are parameter-dependent or multiscale, in the sense that the scale at which points are connected in the cell complex is given by the value $\epsilon$ of the filtration parameter: thus, at an $\epsilon = 0$-scale, all points are disconnected, and they start connecting as the balls grow and intersect.

 The PH aim is to describe how the homologies of a cell complex so constructed are affected by the value of this parameter. BraMAH is, on the contrary, filtration-free and parameterless, in agreement with its aim, which is unambiguously identifying a branched manifold's topology from a point cloud. 

\section*{References}
\bibliography{LORA_topology}

\end{document}